\documentclass[conference]{IEEEtran}
\IEEEoverridecommandlockouts
\usepackage{cite}
\usepackage{amsmath,amssymb,amsfonts}
\usepackage{algorithmic}
\usepackage{graphicx}
\usepackage{textcomp}
\usepackage{xcolor}
\def\BibTeX{{\rm B\kern-.05em{\sc i\kern-.025em b}\kern-.08em
    T\kern-.1667em\lower.7ex\hbox{E}\kern-.125emX}}

\usepackage[utf8]{inputenc}
\usepackage{hyperref}
\usepackage{url}
\usepackage[neverdecrease]{paralist}
\usepackage[caption=false]{subfig}
\usepackage{graphicx}
\usepackage{booktabs}
\usepackage{graphics}
\usepackage{tabularx}
\usepackage{xspace} 
\usepackage{graphicx}
\usepackage{graphicx} 
\usepackage{booktabs}
\usepackage{mdwlist}
\usepackage{multirow}
\usepackage{amsmath}
\usepackage{enumitem} 
\usepackage{calc}
\usepackage{colortbl}
\usepackage[neverdecrease]{paralist}
\usepackage[ruled,linesnumbered]{algorithm2e}
\usepackage{booktabs}
\usepackage{mathtools}
\usepackage{pifont}
\usepackage{color,soul}
\usepackage{array}

\usepackage{tikz}
\usetikzlibrary{shapes.geometric, arrows}

\usepackage{graphicx}

\usepackage{hyperref}

\title{ConSTR: A Contextual Search Term Recommender}

\tikzstyle{startstop} = [rectangle, rounded corners, minimum width=3cm, minimum height=1cm,text centered, draw=black, fill=red!30]

\tikzstyle{io} = [trapezium, trapezium left angle=70, trapezium right angle=110, minimum width=3cm, minimum height=1cm, text centered, draw=black, fill=blue!30]
\tikzstyle{process} = [rectangle, minimum width=3cm, minimum height=1cm, text centered, draw=black, fill=orange!30]
\tikzstyle{decision} = [diamond, minimum width=3cm, minimum height=1cm, text centered, draw=black, fill=green!30]
\tikzstyle{arrow} = [thick,->,>=stealth]
\tikzstyle{process} = [rectangle, minimum width=3cm, minimum height=1cm, text centered, text width=3cm, draw=black, fill=orange!30]

\begin{document}


\author{
    \IEEEauthorblockN{
        Thomas Krämer\IEEEauthorrefmark{1},
        Zeljko Carevic\IEEEauthorrefmark{1},
        Dwaipayan Roy\IEEEauthorrefmark{2},
        Claus-Peter Klas\IEEEauthorrefmark{1},
        Philipp Mayr\IEEEauthorrefmark{1}
    }
    \IEEEauthorblockA{
        \IEEEauthorrefmark{1}GESIS -- Leibniz Institute for the Social Sciences, Cologne, Germany,
        \IEEEauthorrefmark{2}IISER, Kolkata, India\\
        Email: \IEEEauthorrefmark{1}\{thomas.kraemer, zeljko.carevic, philipp.mayr, claus-peter.klas\}@gesis.org,
        \IEEEauthorrefmark{2}dwaipayan.roy@iiserkol.ac.in
    }
}

\maketitle

\begin{abstract}

In this demo paper, we present \textbf{ConSTR}, a novel \textbf{Con}textual \textbf{S}earch \textbf{T}erm \textbf{R}ecommender that utilises the user's interaction context for search term recommendation and literature retrieval. ConSTR integrates a two-layered recommendation interface: the first layer suggests terms with respect to a user’s current search term, and the second layer suggests terms based on the users' previous search activities (interaction context). For the demonstration, ConSTR is built on the arXiv, an academic repository consisting of 1.8 million documents.
\end{abstract}



\section{Introduction}

Exploratory search tasks~\cite{marchionini2006exploratory} that go beyond traditional lookup tasks such as known-item searches usually involve phases of orientation and learning. In particular, the early stages during the information-seeking process (ISP) involve a large degree of uncertainty and vagueness \cite{kuhlthau1993principle}. One challenge that users encounter during the early stages of the ISP is a rather poor understanding of the underlying terminology used in an information system. Furthermore, users are required to provide a concrete representation of an abstract information need in form of a query. Both the lack of knowledge of the underlying terminology and the difficulties in formulating an information need are challenges that users face during complex search tasks. 
To obtain cues about the next steps, users often start with a tentative query and begin exploring the retrieved documents \cite{white2009exploratory}. Particularly the early stage ISP, where a user starts his or her search with such a tentative and vague query, shows a strong need for systemic support. This support is provided by most modern search systems in the form of search term recommenders (STR)~\cite{Mutschke2011,Fuhr2002}. 

\section{ConSTR}
In this paper, we propose \textbf{ConSTR}, a novel \textbf{Con}textual \textbf{S}earch \textbf{T}erm \textbf{R}ecommender (ConSTR) that integrates a two-layer source for recommendations: one layer suggests similar terms based solely on the similarity between the user's current search terms (query) and terms that are similar in an abstract space provided by an embedding model (word2vec, GloVe etc.). The second layer suggests terms on the basis of the user's entire \emph{interaction context}~\cite{Carevic2018}. The interaction context is a representation of the user's previous search activities in the current session, including all prior \textit{queries} and \textit{keywords}. 
By using this two-layer design, we expect that users will benefit from both: immediate recommendations related to the current query (using the first layer), while the use of the interaction context will contain recommendations primarily related to the entire session (by the second layer). 
Our demo application focuses on three goals: $i)$ to recommend adequate search/expansion terms to a given seed query, $ii)$ to make the recommendation process transparent to the user by offering an explicit interface allowing the user to review and adjust the underlying interaction context used for the recommendations, and $iii)$ to enable users to interactively reformulate the queries, getting to know the related term space from the initial -- often vague -- query to a more focused expanded query. 
In particular, our second objective is aimed towards building trust by allowing the user to interact with the recommender system containing interaction context and offering a choice of different models used for recommendation.

STR that support users in finding appropriate query terms are not new and they have been implemented in many modern retrieval systems. However, to the best of our knowledge, an \textit{interactive and transparent STR} that suggests terms to both: the user's current query and the user's interaction context has not been proposed so far.

\section{System Architecture} \label{sec:architecture}
The system is composed of the following components: the preprocessing component uses Python and the search component is implemented as Spring Boot based Web Application backed by an Elasticsearch index of the arXiv open access repository.

\subsection{Building the embedding model}

To train the embedding model, we utilise 
the arXiv dataset\footnote{\url{https://www.kaggle.com/Cornell-University/arxiv}} consisting of 1.8 million articles with the article title, abstract, categories etc. Particularly the abstracts of the papers from the datasets are used for training the embedding model. The skip-gram model of word2vec is used for training the embedding in a 300 dimensional abstract space. The default values of the other parameters are used, and changing those parameters would not change the overall performance as reported in~\cite{RoyGBBM18}.
As another alternative source, we use the pre-trained Wikipedia2014+Gigaword5 embedding model trained using GloVe (in 300 dimensions)\footnote{\url{https://nlp.stanford.edu/projects/glove/}}.
In general, the embedding model in ConSTR can be extended to be used with any embedding model which would provide the flexibility to utilise some field-based recommendation (e.g. medical domain etc.).


\subsection{Keywords and Indexing }
Descriptive metadata such as a document's keywords are a valuable source of information that can be represented in a users interaction context. Keywords, however, are not present in arXiv documents. Hence, we decided to enrich the arXiv index with keywords that we extracted using \em YAKE \em \cite{campos2018yake} - an unsupervised automatic keyword extraction method. We used the abstract of the arXiv documents to extract a maximum of 5 keywords per document with an n-gram size of two.
\subsection{Search application}
\begin{figure}[h!]
\includegraphics[width=\linewidth]{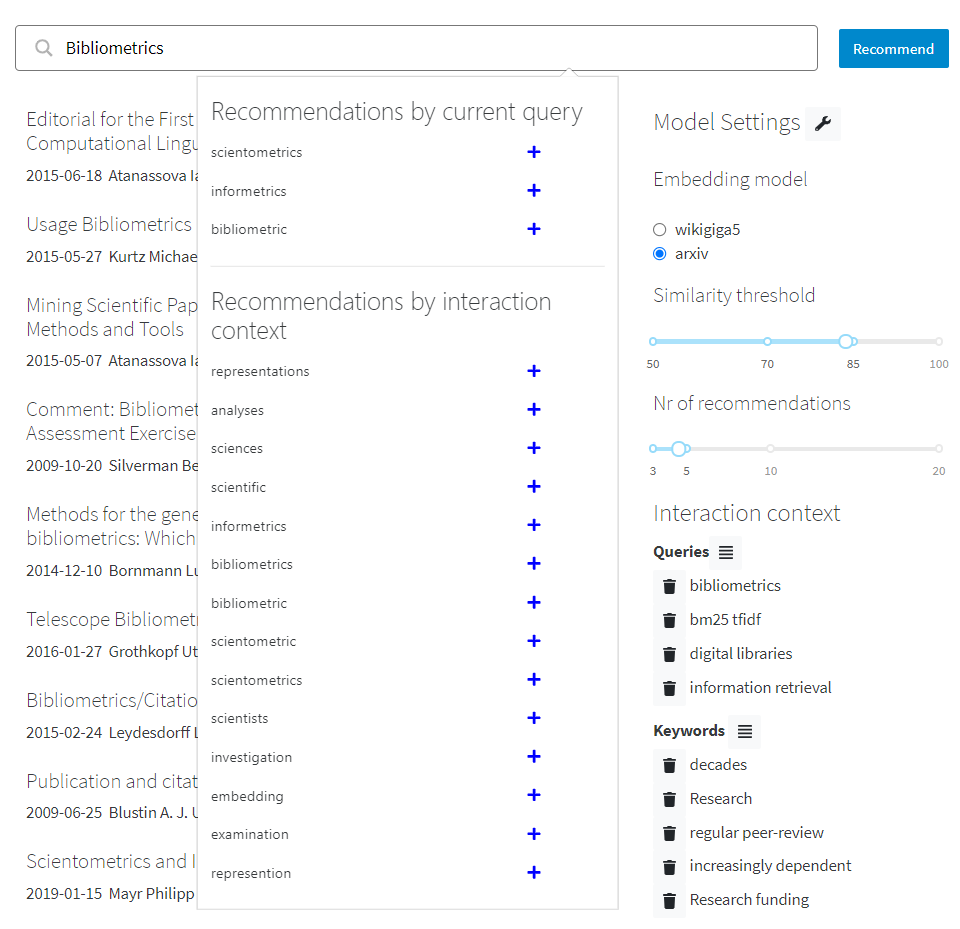}
\caption{ConSTR search view with context, settings and recommendations.}
\end{figure}
The search application is implemented using \href{https://github.com/jhipster/generator-jhipster}{JHipster} and  \href{http://searchkit.github.io/searchkit/stable/}{Searchkit}. 
\subsubsection{Interaction context}
The interaction context consists of queries and keywords. Users can assess past search interactions and remove items from the interaction context with one click. The following interactions are stored in a user's interaction context: 1) each issued query, 2) click on result (keywords of the document) and 3) click on a recommendation (query terms expanded by the recommendations).
\begin{figure}[ht]
\includegraphics[width=\linewidth,height=8cm,keepaspectratio]{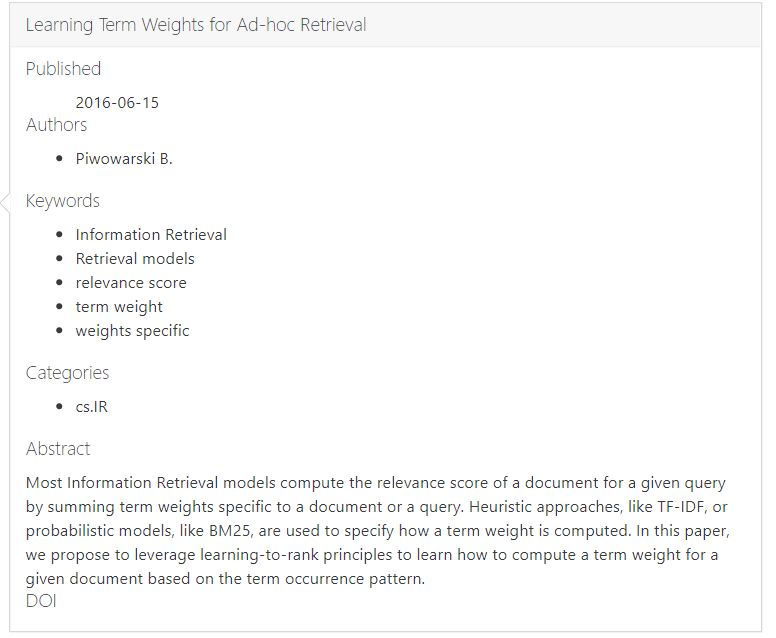}
\caption{ConSTR documents detail pane.}
\end{figure}
\subsubsection{Model settings} 
Users can choose the model (arXiv or Wikipedia+Gigaword5) used to determine the recommended terms.
Users can also specify the similarity threshold (i.e. the minimum similarity score between a seed and a recommended term) to be applied when requesting for the nearest terms to the query, and the number of recommendation.
\subsubsection{Recommendation pane} 
Recommendations are updated when a user issues a search and retrieved for both the current query and the interaction context. They are displayed on the upper (query based) or lower part (interaction context based) of the recommendation pane (see Figure 1). A click on a recommendation will expand the current query and results update automatically. 

\section{Future work}
Currently, recommendations are retrieved for each term of a query. We intend to enable support for recommendations based on phrase queries as well as post-processing (e.g. stemming) before presenting recommendations. The interactions persisted in the interaction context will be extended and include other user actions, such as mouse position or the selection of search facets. Items in the interaction context are ranked and have a timestamp. As the interaction context grows over time, we will implement algorithms taking into account the interaction history and use only selected queries and keywords instead of the entire interaction context.  
As part of future work, ConSTR will be evaluated with real users.

\section{Demo access} 
The demo is available at \href{https://constr.memotaxis.org}{\textbf{constr.memotaxis.org}}.
Further information can be found in the  \href{https://www.youtube.com/playlist?list=PLK4WlSr348zkog2wFGXRaKBxNz7LB5iuP}{\textit{ConSTR demo video}}.



\bibliographystyle{IEEEtran}
\bibliography{IEEEabrv, ref}
\end{document}